\documentclass[dvips,10pt,aps,prc,preprint,groupedaddress,superscriptaddress,showpacs]{revtex4-1}
\usepackage{natbib}
\usepackage{graphicx,subfigure,float,graphics}
\usepackage{epsfig}

\begin{document}

\title{Analytical solution for the Davydov-Chaban Hamiltonian \\ with sextic potential for $\gamma=30^{\circ}$}

\author{P. Buganu}
\email[]{buganu$_$p@yahoo.com}
\affiliation{Department of Theoretical Physics, National Institute for Physics and Nuclear Engineering,\\
Str. Reactorului 30, RO-077125, POB-MG6, Bucharest-M\v{a}gurele, Romania}
\author{R. Budaca}
\affiliation{Department of Theoretical Physics, National Institute for Physics and Nuclear Engineering,\\
Str. Reactorului 30, RO-077125, POB-MG6, Bucharest-M\v{a}gurele, Romania}

\date{\today}

\begin{abstract}
An analytical solution for the Davydov-Chaban Hamiltonian with a sextic oscillator potential for the variable $\beta$ and $\gamma$ fixed to $30^{\circ}$, is proposed. The model is conventionally called Z(4)-Sextic. For the considered potential shapes the solution is exact for the ground and $\beta$ bands, while for the $\gamma$ band an approximation is adopted. Due to the scaling property of the problem the energy and $B(E2)$ transition ratios depend on a single parameter apart from an integer number which limits the number of allowed states. For certain constraints imposed on the free parameter, which lead to simpler special potentials, the energy and $B(E2)$ transition ratios are parameter independent. The energy spectra of the ground and first $\beta$ and $\gamma$ bands as well as the corresponding $B(E2)$ transitions, determined with Z(4)-Sextic, are studied as function of the free parameter and presented in detail for the special cases. Numerical applications are done for the $^{128,130,132}$Xe and $^{192,194,196}$Pt isotopes, revealing a qualitative agreement with experiment and a phase transition in Xe isotopes.
\end{abstract}

\maketitle

\section{\label{sec:intro}Introduction}
Soon after the Bohr-Mottelson Model (BMM) \cite{Bo1,Bo2} was proposed for nuclear structure together with its first solution \cite{Bo1} for spherical nuclei, many attempts were done to improve and extend it by taking into account axial and non axial deformation, coupling between $\beta$ and $\gamma$ vibrations or various anharmonicities. Most of these approaches were reviewed in Refs.\cite{Fort1,Casten,Cej}. A new phase in the field begun with the studies of the phase transitions by means of the classical limits of Hamiltonians constructed from operators belonging to compact Lie algebras \cite{Gil1,Gil2,Gil3}. This algorithm was further used in \cite{Gino} to associate classical shape variables to the Interacting Boson Model (IBM) \cite{Ar1,Ar2,Ar3,Iac1} and allowed to establish the nature of the quantum phase transition \cite{Diep} between the dynamical symmetries, namely, U(5) (spherical vibrator), O(6) ($\gamma$-unstable) and SU(3) (axial rotor). The start of a long series of studies, both theoretical and experimental, was given mainly by two papers in which approximate solutions of the BMM were offered for the critical points of the shape phase transitions U(5)-O(6) and U(5)-SU(3), called E(5) \cite{Iac2} and X(5) \cite{Iac3}, respectively. Other two important models which are worth mentioning here are Y(5) \cite{Iac4} and Z(5) \cite{Bona1} associated with the transitions between the axial and triaxial shapes and respectively between prolate and oblate shapes. The critical point approaches mentioned above have the advantage to be parameter free solutions except for a scaling factor, making them easily verifiable reference points for the experimental data. This is actually a general trait of the exactly solvable models of nuclei \cite{Isac}. Other efforts were also directed to special realisations of the BMM in the view of some constraints imposed on the shape variables or inertial parameters. For example "freezing" the $\gamma$ variable to a certain value in the classical BMM, leads after quantization in curvilinear coordinates to simpler Hamiltonians suitable to describe the special case of the $\gamma$ rigid collective motion. The first study in this direction brought to light the Davydov-Chaban model for rotation-vibration interaction in non-axial nuclei \cite{Dav3}. Later on, an exact solution for this model was proposed \cite{Bona2} in the case of $\gamma=30^{\circ}$, where instead of a displaced harmonic oscillator in $\beta$ shape variable an infinite square well potential was used. The solution called Z(4) due to the similarity to the Z(5) model, inspired other studies of the $\gamma$ rigid solutions \cite{Bona3,Budaca,Budaca1}.

In this paper we propose an analytical solution for the Davydov-Chaban Hamiltonian \cite{Dav3} with $\gamma=30^{\circ}$ and a sextic potential for the only shape variable, i.e. $\beta$. The model is conventionally called Z(4)-Sextic. In this framework, the separation of the angular variables from the $\beta$ shape variable is exact. The differential equation involving Euler angles is solved as in Ref.\cite{Mey}, while that for $\beta$ is brought to a Schr\"{o}dinger form with a sextic potential and a centrifugal-like term. The problem of the sextic potential is not an exactly solvable one because its spectral problem is reduced to the diagonalization of an infinite-dimensional Hamiltonian matrix. However, for a family of potentials whose coefficients satisfy certain relations between them and the factor of the centrifugal term, the problem becomes quasi-exactly solvable \cite{Tur,Ush}, i.e. its infinite Hamiltonian matrix acquire a block diagonal form allowing thus an algebraic treatment for a finite subset of eigenstates. For a physically meaning description, the above mentioned constraints must be corroborated also with the condition of constant potential. Despite these restrictions, the Z(4)-Sextic eigenvalue problem is exactly solved for the ground and $\beta$ bands. Concerning the $\gamma$ band, an approximation is involved in the centrifugal term in order to accommodate all the above restrictions. Due to the scaling property of the exactly solvable sextic potential with an associated centrifugal term, the energy and the $B(E2)$ transitions depend on a single parameter up to an overall scaling factor. Moreover, for particularly interesting shapes of the potential, parameter free expressions are possible for the normalized energies and $B(E2)$ transition probabilities.

The use of such an involved potential is supported by the fact that it is the simplest shape which through continuous variation of its parameters can have either a spherical minimum, a deformed minimum or both. It is worth to mention that exact \cite{Lev1,Lev2} and approximate \cite{Rad1,Rad2} solutions by using a sextic potential were also given, in five dimensions, for E(5) and respectively X(5) and Z(5) related approaches. Other solutions in the vicinity of $\gamma=30^{\circ}$, but with $\gamma$ soft can be found in Refs.\cite{Fort2,Fort3,Baer,Yigi,Inci}.

The present work has the following plan. The Z(4)-Sextic model Hamiltonian is presented in Section II, while its associated $\beta$ differential equation is treated in Section III. In Section IV, one gives the model wave functions and calculate the $B(E2)$ transition probabilities. Extensive numerical results and few model fits to experimental data are given in Section V. The main conclusions are drawn in Section VI.

\section{\label{sec:model}The model Hamiltonian}
\renewcommand{\theequation}{2.\arabic{equation}}

The eigenvalue problem, when the nucleus is $\gamma-$rigid, has the following form \cite{Dav3}:
\begin{equation}
-\frac{\hbar^{2}}{2B}\left[\frac{1}{\beta^{3}}\frac{\partial}{\partial\beta}\beta^{3}\frac{\partial}{\partial\beta}-\frac{1}{4\beta^{2}}\sum_{k=1}^{3}\frac{\hat{Q}_{k}^{2}}{\sin^{2}\left(\gamma-\frac{2\pi}{3}k\right)}\right]\Psi(\beta,\Omega)+V(\beta)\Psi(\beta,\Omega)=E\Psi(\beta,\Omega),
\label{Hi}
\end{equation}
where $B$ is the mass parameter, $\beta$, $\gamma$ and $\hat{Q}_k$ are the intrinsic deformation coordinates and respectively the operators of the total angular momentum projections in the intrinsic reference frame, while with $\Omega$ are denoted the rotation Euler angles $(\theta_1,\theta_2,\theta_3)$. Here, $\gamma$ is considered a parameter and not a variable, such that when the kinetic energy of the classical BMM is quantized in curvilinear coordinates one arrives at the Hamiltonian (\ref{Hi}) which depends only on four variables $(\beta,\Omega)$. When $\gamma=\pi/6$, two moments of inertia in the intrinsic reference frame become equal and then the rotational term reads:
\begin{equation}
\frac{1}{4}\sum_{k=1}^{3}\frac{\hat{Q}_{k}^{2}}{\sin^{2}\left(\gamma-\frac{2\pi}{3}k\right)}=(\hat{Q}^{2}-\frac{3}{4}\hat{Q}_{1}^{2}).
\end{equation}
The separation of variables is achieved by considering the wave function $\Psi(\beta,\Omega)=\phi(\beta)\psi(\Omega)$ which leads to the following equation in $\beta$ variable:
\begin{equation}
\left[-\frac{1}{\beta^{3}}\frac{d}{d\beta}\beta^{3}\frac{d}{d\beta}+\frac{W}{\beta^{2}}+v(\beta)\right]\phi(\beta)=\varepsilon\phi(\beta),
\label{eqbeta}
\end{equation}
where the following notations are used $v(\beta)=\frac{2B}{\hbar^{2}}V(\beta)$ and $\varepsilon=\frac{2B}{\hbar^{2}}E$, while $W$ is the eigenvalue for the equation of the angular part,
\begin{equation}
\left(\hat{Q}^{2}-\frac{3}{4}\hat{Q}_{1}^{2}\right)\psi(\Omega)=W\psi(\Omega).
\label{eqomega}
\end{equation}
The above equation was solved in Ref.\cite{Mey} with the results:
\begin{equation}
W=W_{LR}=L(L+1)-\frac{3}{4}R^{2},
\end{equation}
and
\begin{equation}
\psi_{\mu R}^{L}(\Omega)=\sqrt{\frac{2L+1}{16\pi^{2}(1+\delta_{R,0})}}\left[D_{\mu,R}^{(L)}(\Omega)+(-1)^{L}D_{\mu,-R}^{(L)}(\Omega)\right].
\label{ang}
\end{equation}
Here $D_{\mu,R}^{(L)}(\Omega)$ are the Wigner functions associated to the total angular momentum $L$ and its projections on the body fixed x-axis and laboratory fixed z-axis, $R$ and respectively $\mu$. For the energy spectrum it is more advantageous to use instead of $R$ the wobbling quantum number $n_{\omega}=L-R$ which for the ground and $\beta$ bands is $n_{\omega}=0$, while for the $\gamma$ band it takes the values $n_{\omega}=1$ for $L$ odd and $n_{\omega}=2$ for $L$ even. Within this convention the eigenvalue of the angular part of the problem is written as
\begin{equation}
W_{LR}=W_{Ln_{w}}=L(L+1)-\frac{3}{4}(L-n_{w})^{2}.
\end{equation}

\section{Solution for the $\beta$ part of the Hamiltonian}
\renewcommand{\theequation}{3.\arabic{equation}}

It is convenient to bring Eq. (\ref{eqbeta}) to a Schr\"{o}dinger form. This is realized by changing the function with $\phi(\beta)=\beta^{-\frac{3}{2}}\varphi(\beta)$ \cite{Dav3}:
\begin{equation}
\left[-\frac{d^{2}}{d\beta^{2}}+\frac{W_{Ln_{w}}+\frac{3}{4}}{\beta^{2}}+v(\beta)\right]\varphi(\beta)=\varepsilon\varphi(\beta).
\label{ecschr}
\end{equation}
Further, Eq. (\ref{ecschr}) is compared with the exactly solvable case of the sextic potential \cite{Ush} which leads to the following correspondences:
\begin{equation}
W_{Ln_{w}}+\frac{3}{4}=\left(2s-\frac{1}{2}\right)\left(2s-\frac{3}{2}\right),
\end{equation}
\begin{equation}
v(\beta)=\left[b^{2}-4a\left(s+\frac{1}{2}+M\right)\right]\beta^{2}+2ab\beta^{4}+a^{2}\beta^{6}.
\label{vbeta}
\end{equation}
The potential (\ref{vbeta}) depends on two parameters, $a$ and $b$, and on $L$ and $n_{w}$ quantum numbers through $s$. $M$ is a natural number which establishes the number of states that can be determined. This implication will be explained later when discussing the wave functions. The number of parameters is reduced to a single one by changing the variable with $\beta=ya^{-\frac{1}{4}}$. Then by introducing the notations $\alpha=\frac{b}{\sqrt{a}}$ and $\varepsilon_{y}=\frac{\varepsilon}{\sqrt{a}}$ one gets:
\begin{equation}
\left[-\frac{d^{2}}{dy^{2}}+\frac{W_{Ln_{w}}+\frac{3}{4}}{y^{2}}+(\alpha^{2}-4c)y^{2}+2\alpha y^{4}+y^{6}\right]\eta(y)=\varepsilon_{y}\eta(y),
\label{eqy}
\end{equation}
where
\begin{equation}
c\equiv s+\frac{1}{2}+M.
\end{equation}
Because $s$ depends on $L$ and $n_{w}$, the potential of Eq. (\ref{eqy}) is state dependent. For the ground and $\beta$ bands $n_{w}=0$, such that
\begin{equation}
s=\frac{L}{4}+1,\;\;c=M+\frac{L}{4}+\frac{3}{2}.
\label{sc}
\end{equation}
In order to have a state invariant potential for this case of ground and $\beta$ band states, the following condition must be satisfied:
\begin{equation}
c=M+\frac{L}{4}+\frac{3}{2}=const.
\label{cond}
\end{equation}
It can be easily checked that the above condition is satisfied if $M$ is decreased with one unit when $L$ is increased with four. This means that for $L/2$ even and $L/2$ odd there are two different constants $c$:
\begin{eqnarray}
(M,L)&:&(K,0),(K-1,4),...\Rightarrow K+\frac{3}{2}=c^{K}_0,\\
(M,L)&:&(K,2),(K-1,6),...\Rightarrow K+2=c^{K}_2,
\label{rule1}
\end{eqnarray}
which differ from each other just by $1/2$. Note that the value of $K=M_{max}$ puts a limit on the number of states which might be determined. For example if $K=1$, the maximum angular momentum state which could be analytically described would be the $L=6$ state, while for $K=2$, the $L=10$ state and so on. This is actually a direct consequence of the condition (\ref{cond}). In case of the $\gamma$ band, when $n_{w}=1$ and 2, $s$ becomes irrational such that the Eq.(\ref{eqy}) cannot be solved anymore for $M$ integer and with constant potential condition fulfilled. A possible way to handle this problem is to extract from the centrifugal term the quantities $3(L-1/2)/2y^2$ and $3(L-1)/y^2$ for $L$ odd and respectively $L$ even, and to replace $y^2$ with its average $\langle y^2\rangle$ on $\eta(y)$ eigenstates of the remaining Hamiltonian for each angular momentum $L$. With these approximations, $s$ and $c$  from the $\gamma$ band will have the same expressions (\ref{sc}) as in the case of ground and $\beta$ bands. Moreover, for $L$ even states of the $\gamma$ band, $c$ will have the same two values as for the ground and $\beta$ bands. While for $L$ odd, $c$ will have other two values:
\begin{eqnarray}
(M,L):(K,1),(K-1,5),...\Rightarrow K+\frac{7}{4}&=&c^{K}_1,\\
(M,L):(K,3),(K-1,7),...\Rightarrow K+\frac{9}{4}&=&c^{K}_3,
\label{rule2}
\end{eqnarray}
corresponding to $(L-1)/2$ even and respectively $(L-1)/2$ odd. Finally, the four values of the constant $c$ can be summarized by the formula:
\begin{equation}
c_{m}^{K}=c_{0}^{K}+\frac{1}{4}m=K+\frac{3}{2}+\frac{1}{4}m,\,\,\,m=0,1,2,3.
\end{equation}
The condition of the constant potential is then exactly satisfied for four distinct sets of states, which correspond to slightly different potentials. This picture is improved by considering for the general potential the following form:
\begin{equation}
v_{m}^{K}(y)=(\alpha^{2}-4c^{K}_m)y^{2}+2\alpha y^{4}+y^{6}+u^{K}_m(\alpha).
\label{poty}
\end{equation}
For a fixed $K$, $u_m^{K}$ are constants depending on $\alpha$, which are fixed such that the minimum energy of the potentials $v_m^{K}$ to be the same. Choosing $u_{0}^{K}=0$, the other constants are given by:
\begin{equation}
u_{i}^{K}=\left(\alpha^{2}-4c_0^{K}\right)\left(y_{0,0}^{K}\right)^{2}-\left(\alpha^{2}-4c_i^{K}\right)\left(y_{0,i}^{K}\right)^{2}+2\alpha\left[\left(y_{0,0}^{K}\right)^{4}-\left(y_{0,i}^{K}\right)^{4}\right]+\left(y_{0,0}^{K}\right)^{6}-\left(y_{0,i}^{K}\right)^{6},\nonumber
\end{equation}
with $i=1,2,3$ and $y_{0,m}^{K}$ being the minimum points:
\begin{equation}
(y_{0,m}^{K})^{2}=\frac{1}{3}(-2\alpha\pm\sqrt{\alpha^{2}+12c_m^{K}}).
\end{equation}
Taking the ansatz function \cite{Ush}
\begin{equation}
\eta_{M}(y)\sim P_{M}(y^2)y^{2s-\frac{1}{2}}e^{-\frac{y^4}{4}-\frac{\alpha y^{2}}{2}},
\end{equation}
the Eq. (\ref{eqy}) is then reduced to the following differential equation,
\begin{equation}
\left[-\left(\frac{d^{2}}{dy^{2}}+\frac{4s-1}{y}\frac{d}{dy}\right)+2\alpha y\frac{d}{dy}+2y^{2}\left(y\frac{d}{dy}-2M\right)\right]P_{M}(y^{2})=\lambda P_{M}(y^{2}),
\label{eqQ}
\end{equation}
where $P_{M}(y^{2})$ are polynomials in $y^2$ of order $M$. The eigenvalues $\lambda$ are obtained for each $M$ using the analytical procedure given in Appendix of Ref.\cite{Rad2}. For each value of $M$ there are $M+1$ solutions which are differentiated by the $\beta$ vibrational quantum number $n_{\beta}$ in the following way: The lowest eigenvalue $\lambda$ corresponds to $n_{\beta}=0$, while the highest to $n_{\beta}=M+1$. For the present physical problem only the solutions with $n_\beta=0$ and $n_\beta=1$ will be considered, which correspond to the ground and $\gamma$ bands and respectively to the $\beta$ band states. $\lambda$ also depends on $L$ through $s$ and one must remind that at this point $L$ and $M$ are interdependent through the condition (\ref{cond}), the actual relationship being dictated by the value of $K$. Thus, the $M$ indexing of $\lambda$ will be replaced from here by $K$. Following all the algebraic manipulations which lead to Eq.(\ref{eqQ}) and taking into account the above considerations, $\lambda$ can be alternatively expressed as:
\begin{equation}
\lambda=\lambda_{n_{\beta}L}^{K}=\varepsilon_{y}-u_{m}^{K}-4\alpha s-\frac{\frac{3}{2}\left(L-\frac{1}{2}\right)}{\langle y^{2}\rangle_{KL}}\delta_{n_{\omega},1}-\frac{3(L-1)}{\langle y^{2}\rangle_{KL}}\delta_{n_{\omega},2}.
\end{equation}
From the above relation one finally extracts the total energy of the system:
\begin{equation}
E_{n_{\beta},n_{\omega},L}=\frac{\hbar^{2}\sqrt{a}}{2B}\left[\lambda_{n_{\beta}L}^{K}+4\alpha \left(\frac{L}{4}+1\right)+u_{m}^{K}+\frac{\frac{3}{2}\left(L-\frac{1}{2}\right)}{\langle y^{2}\rangle_{KL}}\delta_{n_{\omega},1}+\frac{3(L-1)}{\langle y^{2}\rangle_{KL}}\delta_{n_{\omega},2}\right],
\label{ener}
\end{equation}
which is indexed by the $\beta$ vibration and wobbling quantum numbers $n_{\beta}$ and $n_{w}$, as well as by the intrinsic angular momentum $L$. The index $m$ is completely determined only by $L$. Although the above energy also depends on the integer $K$, this is not a true quantum number but rather a special kind of parameter. Similarly to the eigenvalue $\lambda$, the associated eigenfunctions of Eq.(\ref{eqQ}) also depend on $K$ and $L$. Such that due to the orthogonality of the angular wave functions (\ref{ang}), the average of $y^{2}$ entering in the definition of the total energy are only $K$ and $L$ dependent. From Eq.(\ref{ener}) one can see that the energy spectrum normalized to the energy of the first excited state depends only on the parameter $\alpha$ and the integer $K$. For further calculations one defines the energy ratios:
\begin{equation}
R(n_{\beta},n_{\omega},L,\alpha)=\frac{E_{n_{\beta},n_{\omega},L}-E_{0,0,0}}{E_{0,0,2}-E_{0,0,0}},
\label{enrat}
\end{equation}
for a fixed value of $K$.

\section{Total wave functions and $B(E2)$ transition rates}
\renewcommand{\theequation}{4.\arabic{equation}}

As was explained in Section II, the total wave function is factorized into an angular part and a $\beta$ depending factor function:
\begin{equation}
\Psi_{n_{\beta}LR}^{M}(\beta,\Omega)=\psi_{\mu R}^{L}(\Omega)\phi_{n_{\beta}L}^{M}(\beta),
\end{equation}
where the angular factor state was defined by Eq.(\ref{ang}) keeping the notation with $R$ instead of $n_{w}$ for convenience in calculating angular matrix elements. In what concerns the $\beta$ wave function, it has the following form:
\begin{equation}
\phi_{n_{\beta}L}^{M}(\beta)=\sqrt{a}\phi_{n_{\beta}L}^{M}(y)=\sqrt{a}N_{n_{\beta}L}^{M}(\alpha)P_{Mn_{\beta}}(y^2)y^{\frac{L}{2}}e^{-\frac{y^4}{4}-\frac{\alpha y^{2}}{2}},
\end{equation}
with $y=\beta a^{1/4}$ and $N_{n_{\beta}L}^{M}(\alpha)$ being the normalization constant with respect to the $y^{3}dy$ integration measure. As was already mentioned when the expression of the total energy was discussed, $M$ is uniquely determined by $L$ for a fixed value of $K$. Thus, a more natural dependence of the total wave function would be on $K$ instead of $M$. However $M$ express more intuitively the analytical form of the $\beta$ factor state.

\begin{figure}[th!]
\begin{center}
\includegraphics[width=0.7\textwidth]{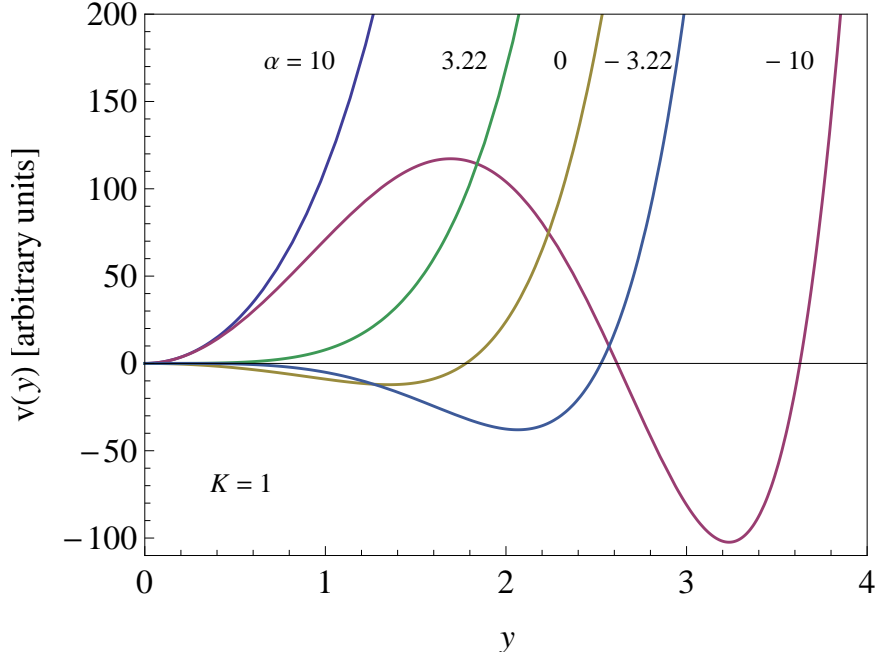}
\end{center}
\caption{The shapes of the energy potential for $\gamma$ fixed at $\pi/6$ and with $K=1$, corresponding to $\alpha=-10$, $\pm2\sqrt{c_0} (\pm3.22)$, $0$, $10$ are plotted as function of $y$.}
\label{fig1}
\end{figure}

Having the analytical expression of the total wave function, one can readily compute the $B(E2)$ transition probabilities. The quadrupole operator for Z(4)-Sextic has the same form as for the Z(4) solution \cite{Bona2},
\begin{equation}
T_{\mu}^{(E2)}=-\frac{1}{\sqrt{2}}t\beta\left(D_{\mu,2}^{(2)}(\Omega)+D_{\mu,-2}^{(2)}(\Omega)\right).
\label{optran}
\end{equation}

The reduced $E2$ transition probabilities are defined as:
\begin{equation}
B(E2,L_{i}\rightarrow L_{f})=|\langle\Psi_{n_{\beta i}L_{i}}^{M}||T_{2}^{(E2)}||\Psi_{n_{\beta f}L_{f}}^{M}\rangle|^{2},
\label{tran}
\end{equation}
where the Rose's convention \cite{Rose} was used for the reduced matrix elements. The matrix elements over the $\beta$ can be rewritten in terms of $y$ with the following result:
\begin{equation}
\langle\phi_{n_{\beta i}L_{i}}^{M}(\beta)|\beta|\phi_{n_{\beta f}L_{f}}^{M}(\beta)\rangle=a^{-\frac{1}{4}}\int_{0}^{\infty}\phi_{n_{\beta i}L_{i}}^{M}(y)y\phi_{n_{\beta f}L_{f}}^{M}(y)y^{3}dy.
\end{equation}

\section{Numerical results}
\renewcommand{\theequation}{5.\arabic{equation}}

As its construction is suggesting, the Z(4)-Sextic model, introduced in the previous sections, is adequate for the description of triaxial nuclei having a $\gamma$ rigidity of $30^{\circ}$. The model depends on a single parameter $\alpha$, apart from a scaling factor and the integer number $K$ which gives the extension of the exactly solvable subspace. Depending on the free parameter $\alpha$ and regardless of the $K$ value, the sextic potential (\ref{poty}) may have a spherical minimum $(\alpha>2\sqrt{c_3^{K}})$, a deformed one $(-2\sqrt{c_0^{K}}<\alpha<2\sqrt{c_0^{K}})$ and simultaneously spherical and deformed minima $(\alpha<-2\sqrt{c_3^{K}})$ for all considered states. These situations are depicted in Fig. \ref{fig1} for $K=1$ where one also showed the particular cases of $\alpha=0$, $\alpha=-2\sqrt{c_0^{1}}$ and $\alpha=2\sqrt{c_0^{1}}$ with the latter one corresponding to a potential shape close to that of the infinite square well. The different constants $c_{m}^{K}$ used for the four distinct sets of states depending on the parity of $L/2$ or $(L-1)/2$, define some small extension intervals where few of the above solutions coexists for different states. If all the constants $c_{m}^{K}$ would be equal, the coexisting intervals would shrink to a single point value of $\alpha$.

\begin{figure*}[hp!]
\begin{center}
\includegraphics[width=0.89\textwidth]{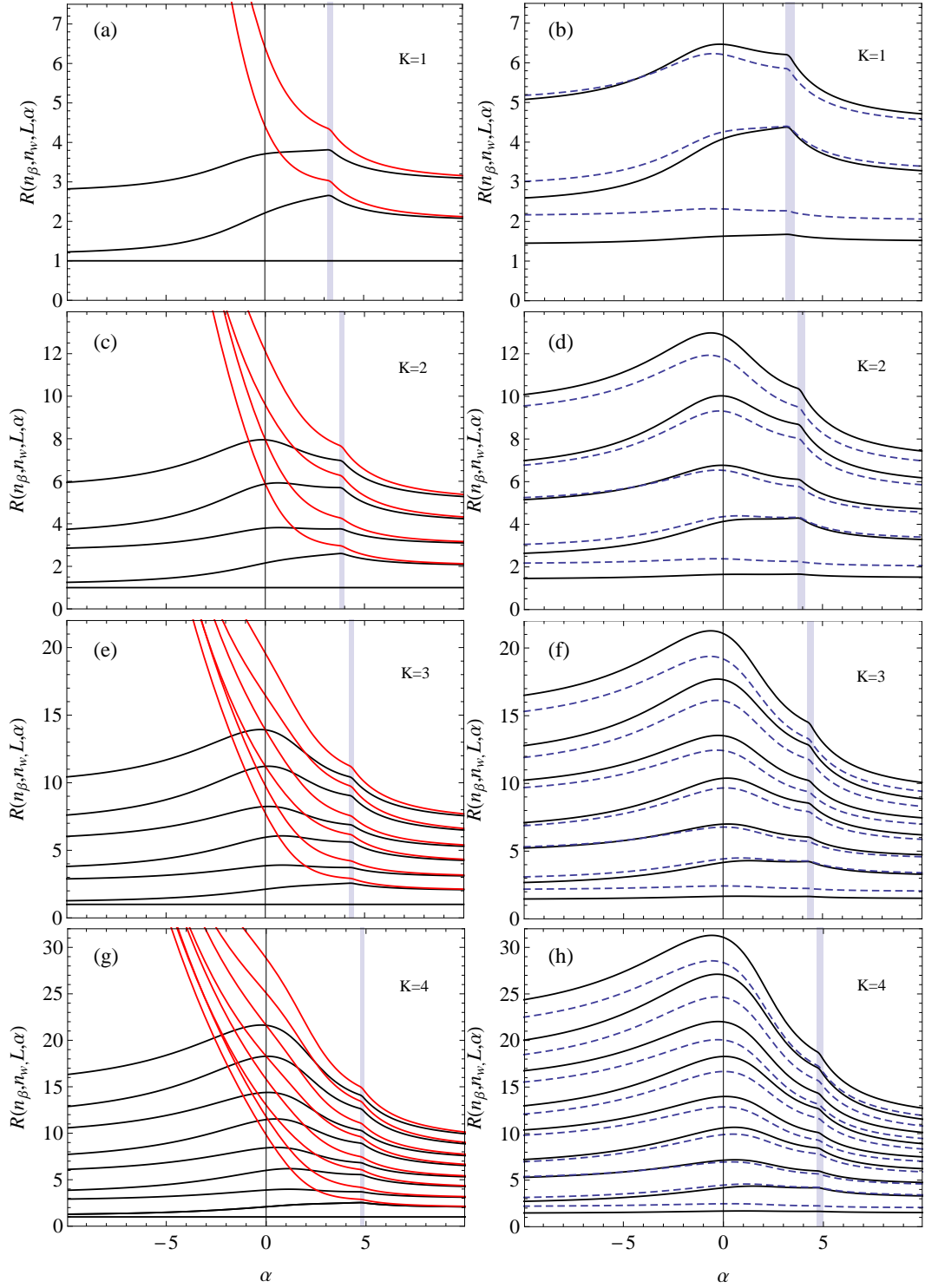}
\vspace{-1.5cm}
\end{center}
\caption{The energy spectrum given by Eq. (\ref{enrat}) is shown as function of $\alpha$ in the interval $[-10,10]$ for $K=1,2,3$ and 4. In the panels (a), (c), (e) and (g) are plotted the energy curves of the ground band and $\beta$ band which go to infinity when $\alpha\rightarrow-\infty$. While in (b), (d), (f) and (h) panels are those corresponding to the $\gamma$ band, with the continuous and dashed curves representing $L$ even and $L$ odd states, respectively.}
\label{fig2}
\end{figure*}

\begin{figure*}[ht!]
\begin{center}
\includegraphics[width=0.99\textwidth]{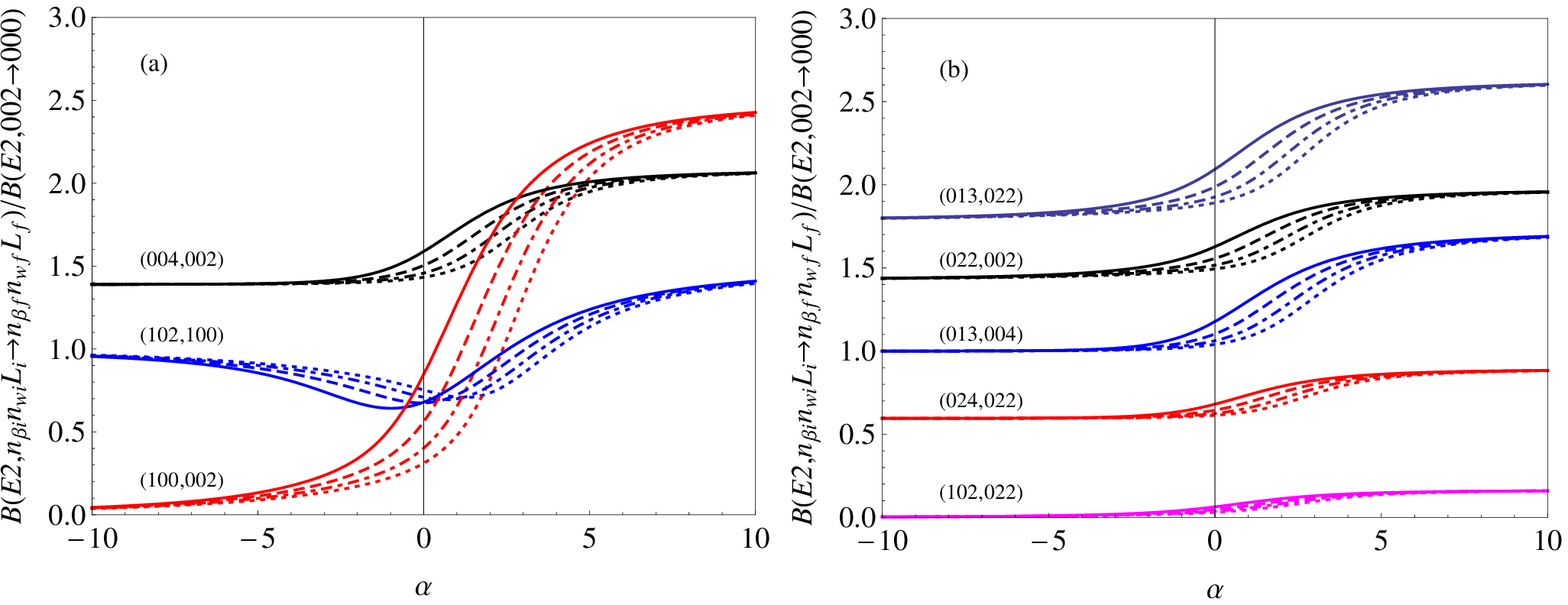}
\end{center}
\caption{The $B(E2)$ transitions $4_{g}^{+}\rightarrow2_g^{+}$, $2_{\beta}^{+}\rightarrow0_{\beta}^{+}$, $0_{\beta}^{+}\rightarrow2_{g}^{+}$ (a) and $3_{\gamma}^{+}\rightarrow2_{\gamma}^{+}$, $2_{\gamma}^{+}\rightarrow2_{g}^{+}$, $3_{\gamma}^{+}\rightarrow4_{g}^{+}$, $4_{\gamma}^{+}\rightarrow2_{\gamma}^{+}$, $2_{\beta}^{+}\rightarrow2_{\gamma}^{+}$ (b), normalized to $B(E2,2_{g}^{+}\rightarrow0_{g}^{+})$ are plotted as function of $\alpha$ in the same interval $[-10,10]$. The continuous, dashed, dot-dashed and dotted lines correspond to $K=1,2,3$ and $4$, respectively. }
\label{fig3}
\end{figure*}

\begin{figure}[ht!]
\begin{center}
\includegraphics[width=0.6\textwidth]{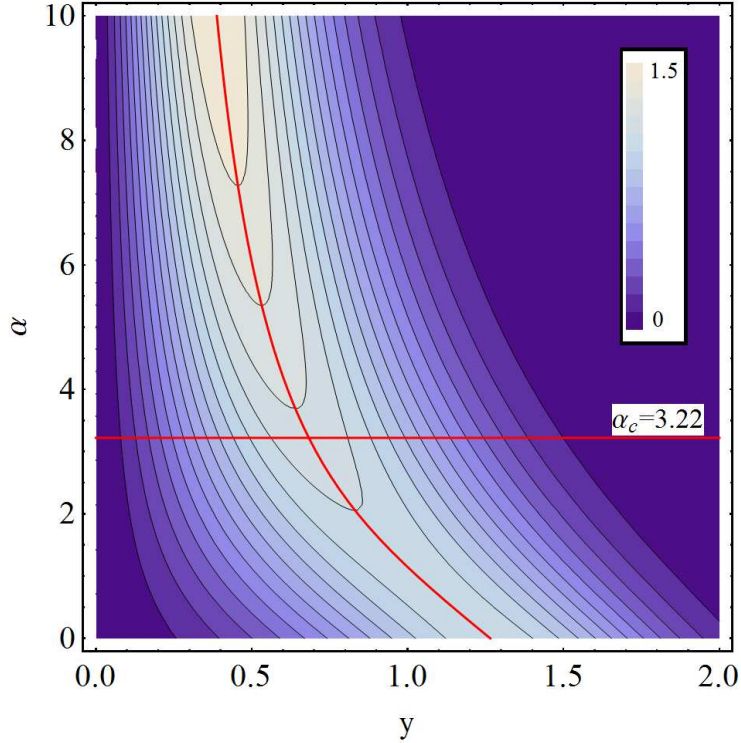}
\end{center}
\caption{The plot of $\phi_{00}^{1}(y)y^{3/2}$ as function of $y$ for $\alpha\in[0,10]$. The square of this quantity is the probability density for the ground state deformation with respect to the $dy$ integration measure. The horizontal red line marks the the wave function behaviour at the critical value $\alpha_{c}$. The evolution of the peak's position with $\alpha$ is visualized by another red curve. The difference between two consecutive contour lines is 0.1.}
\label{wf}
\end{figure}

The advantage of the present model's dependence on a single parameter is that one can study how its characteristics are changed between the pictures discussed above by continuously varying the free parameter. In order to do this and cover all the above mentioned cases, the energy ratios (\ref{enrat}) and few interband and intraband $B(E2)$ transitions (\ref{tran}) normalized to the transition $2_{g}^{+}\rightarrow0_{g}^{+}$ are presented in Fig. \ref{fig2} and Fig. \ref{fig3}, respectively, for a sufficiently large interval of $\alpha$ in order to achieve convergence at both sides. The numerical results visualized in Figs. \ref{fig2} and \ref{fig3} are performed for $K=1,2,3$ and 4. For each $K$ there is a limited number of available states which are exactly determined in the present model. The number of such states in the ground, $\beta$, $\gamma$ with $L$ even and $\gamma$ with $L$ odd bands increases with two when $K$ is increased with one unit. The common parts of the energy spectra corresponding to different $K$ are very similar. However, there are some clear differences, such as the width and the position of the coexistence intervals identified in Fig. \ref{fig2} by the gray area where for a set of states the potential shape has a spherical minimum while for another set it has a deformed minimum. For the ground and $\beta$ bands the interval is $[2\sqrt{c_0^{K}},2\sqrt{c_2^{K}}]$, while for the $\gamma$ band the interval is bigger $[2\sqrt{c_0^{K}},2\sqrt{c_3^{K}}]$. Indeed, as $K$ increases, the grey band becomes thinner and its position moves to higher values of $\alpha$. As a matter of fact in this existence interval, one observes a kink in the energy curves which happens at a critical value $\alpha_{c}$. This value corresponds to the absolute maximum of the signature ratio $R_{4/2}(\alpha)=R(0,0,4,\alpha)$ and is interpreted as the critical point for a first order shape phase transition between spherical and deformed shapes in the framework of presently adopted sextic potential. This is also supported by the fact that the first derivative of the energy in the critical point $\alpha_{c}$ has a discontinuity. As was explained in Ref.\cite{Casten}, the critical point for a first order phase transition corresponds to the situation when the spherical and deformed minima of the potential energy are degenerated. This happens in our case at $\alpha_{c}$ where the potential shape is the flattest, being extended over a wide range of non zero deformations. A similar critical point was pointed out in the analysis made in Ref. \cite{Lev2} regarding a $\gamma$ unstable realization of the sextic potential and in Ref. \cite{Rowe} where a quartic potential was involved. The fact that the critical value $\alpha_{c}$ separates two distinct shape phases can also be seen from the behaviour of the wave function. For example from the dependence of the normalized wave function $\phi_{00}^{1}(y)y^{3/2}$ on $\alpha$ and $y$, shown in Fig. \ref{wf}, one can see that up to the critical value $\alpha_{c}$ the position of the peak changes very rapidly, while for $\alpha>\alpha_{c}$ its position do not vary significantly. This suggests that the two shape phases delimited by $\alpha_{c}$ have different properties.

Contrary to the energy spectra, the $B(E2)$ transition probabilities shown in Fig. \ref{fig3} have a smooth behaviour as function of $\alpha$. While the $K$ variation, induce only a small shift to the right of the curves from Fig. \ref{fig3}. The common feature of the all considered transitions is that their corresponding probabilities become "$K$ degenerate" for $\alpha\rightarrow\pm\infty$ and more sooner for the interband transitions.

\begin{figure*}[ht!]
\begin{center}
\includegraphics[width=0.85\textwidth]{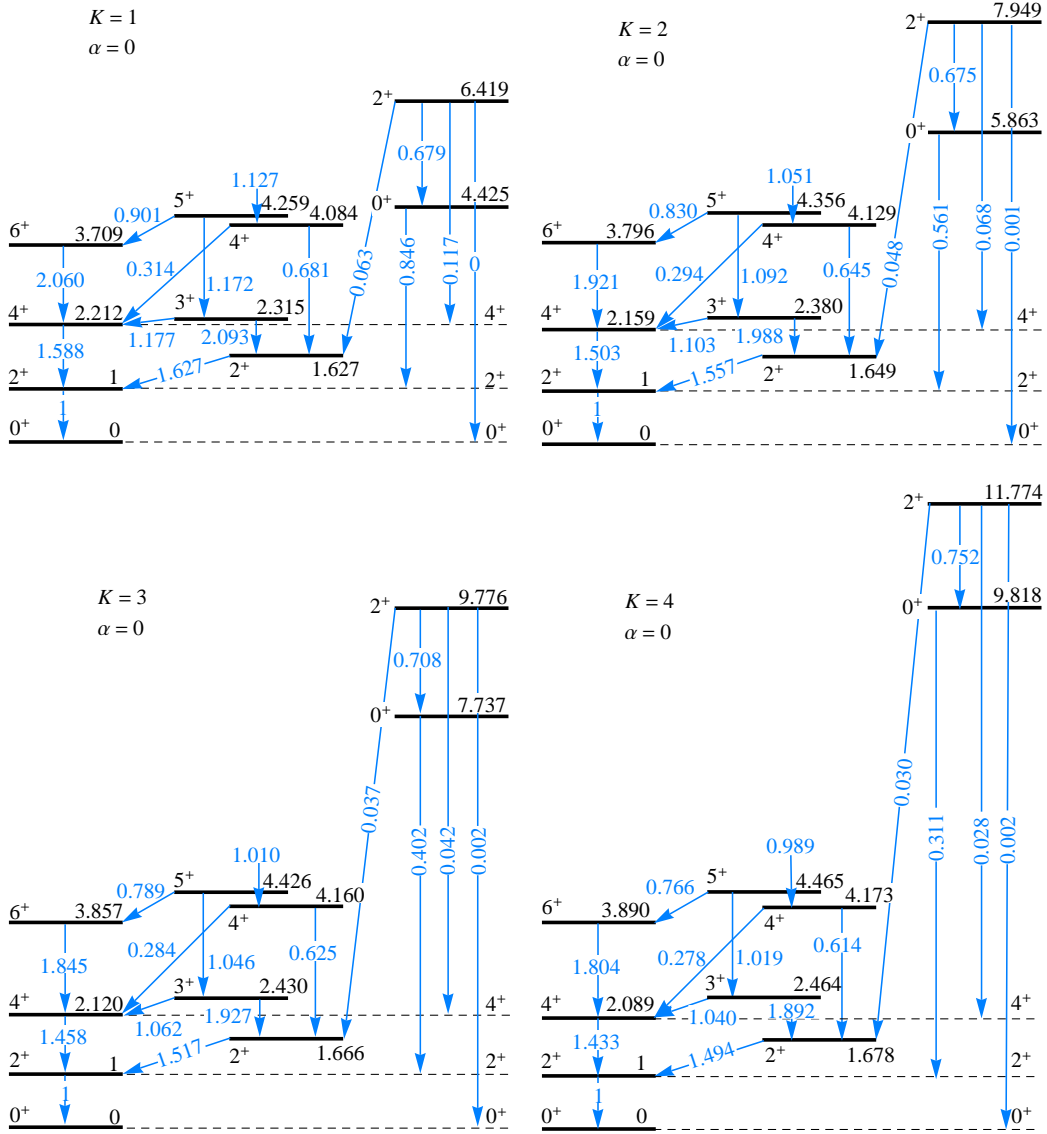}
\end{center}
\caption{Energy spectra and some $E2$ transitions, normalized to the energy of the state $2_{g}^{+}$ and respectively to the transition probability $B(E2,2_{g}^{+}\rightarrow0_{g}^{+})$, are visualized for each $K=1,2,3$ and 4 when $\alpha=0$.}
\label{fig4}
\end{figure*}

\begin{figure*}[ht!]
\begin{center}
\includegraphics[width=0.85\textwidth]{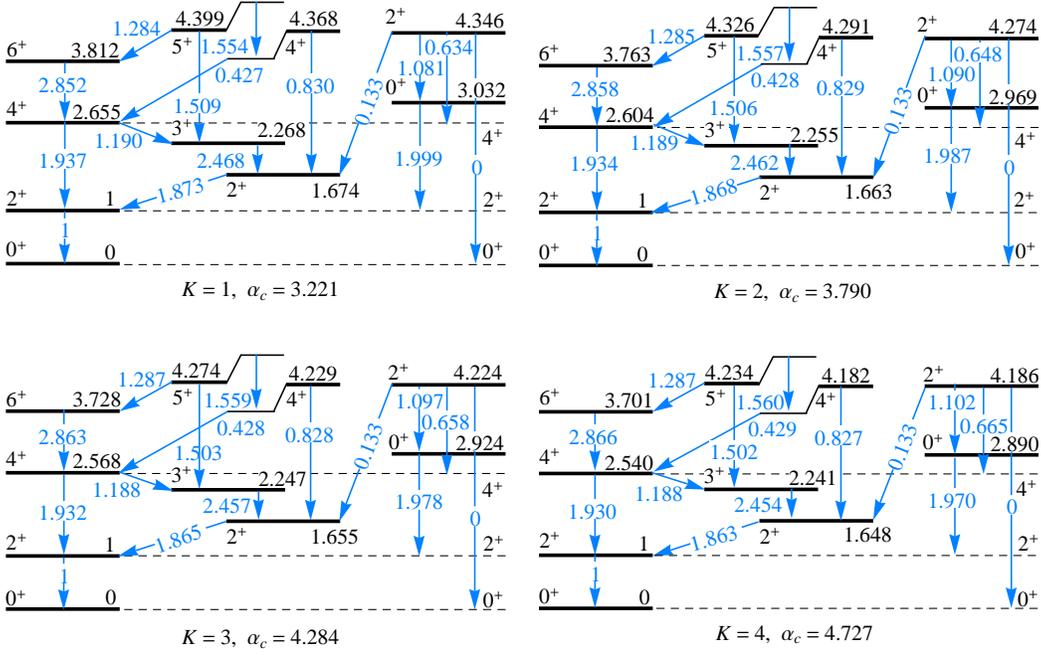}
\end{center}
\caption{The same as in Fig. \ref{fig4}, but for $\alpha=\alpha_{c}$.}
\label{fig5}
\end{figure*}

As was mentioned before, in the coexistence region, and especially at the critical value $\alpha_{c}$, the shape of the potential approximated by $v\approx2\alpha_{c}y^{4}+y^{6}$ is the flattest one, which is consistent with critical point behavior. Moreover, the potential at $\alpha_{c}$ simulates quite well an infinite square well, supported also by the fact that the corresponding energy spectrum is very close to that of Z(4) model. Another interesting aspect of the present model is that some energy ratios curves of the ground and $\beta$ bands are intersecting each other for $\alpha=0$ and becoming thus degenerate. This can be seen only starting from $K=2$, where the last two ground band states $L_{Max}$ and $L_{Max}-2$ are degenerate with $L_{Max}-8$ and $L_{Max}-10$ from the $\beta$ band when $\alpha=0$. This degeneracy may reveal some symmetry properties associated with the resulting simple potential shape $v^{K}_{m}\sim-4c_{m}^{K}y^{2}+y^{6}$. The low lying energy spectrum with a complete set of $E2$ transition probabilities for this special case is graphically represented in Fig. \ref{fig4} for each considered value of $K$. A similar representation is provided in Fig. \ref{fig5} for the other special case corresponding to $\alpha_{c}$ where one also given its numerical value. The parameter free results presented in Figs. \ref{fig4} and \ref{fig5} can be used in a first step as reference points for finding candidate nuclei and then to vary $\alpha$ for a better agreement with the experimental data.

Another important touchstone of the present formalism represents the exact reproduction of the Z(4)-$\beta^{2}$ model \cite{Mac} spectra when $\alpha\rightarrow\infty$. This means that the ground and $\beta$ bands are degenerated and have a harmonic oscillator type spectrum, while the even and odd angular momentum states of the $\gamma$ band deviate from this behavior. In what concerns the other limit, $\alpha\rightarrow-\infty$, the ground and $\gamma$ band spectra achieve a convergence at a noncollective value of $R_{4/2}<2$. While the $\beta$ band energy curves go to infinity. The limiting value $R_{4/2}=2$ is reached at $\alpha=-0.964,-0.804,-0.666$ and $-0.545$ for $K=1,2,3$ and 4 respectively.

\begin{figure*}
\begin{center}
\includegraphics[width=0.73\textwidth]{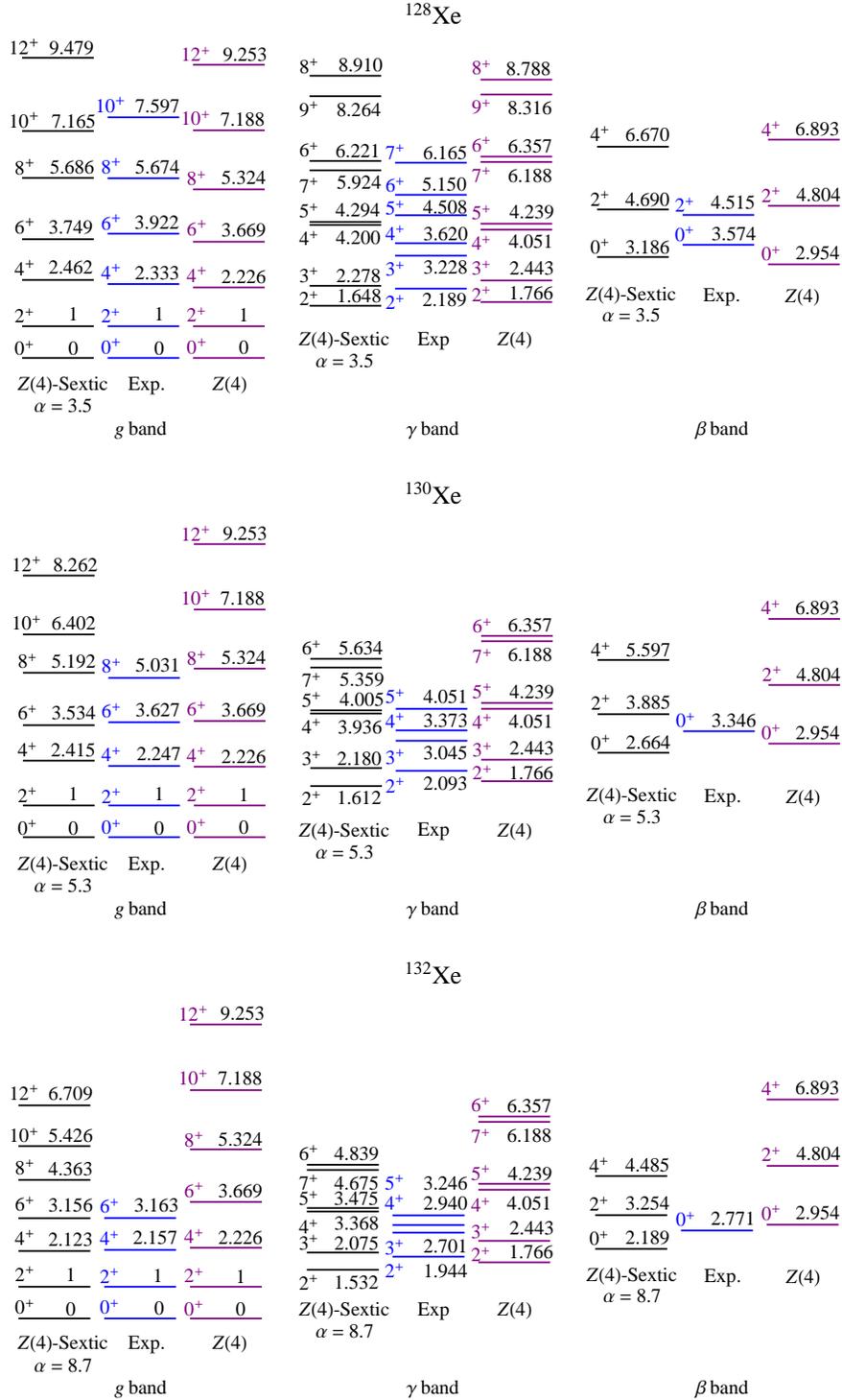}
\end{center}
\caption{The theoretical energy spectra, given by Eq. (\ref{enrat}), are compared with the experimental data \cite{Kan,Bal1,Yu} of the $^{128,130,132}$Xe isotopes and with the Z(4) model predictions. The corresponding rms values for Z(4)-Sextic are 0.516, 0.478 and 0.403, while for Z(4) one obtains 0.528, 0.389 and 0.611.}
\label{fig6}
\end{figure*}

\begin{figure*}
\begin{center}
\includegraphics[width=0.73\textwidth]{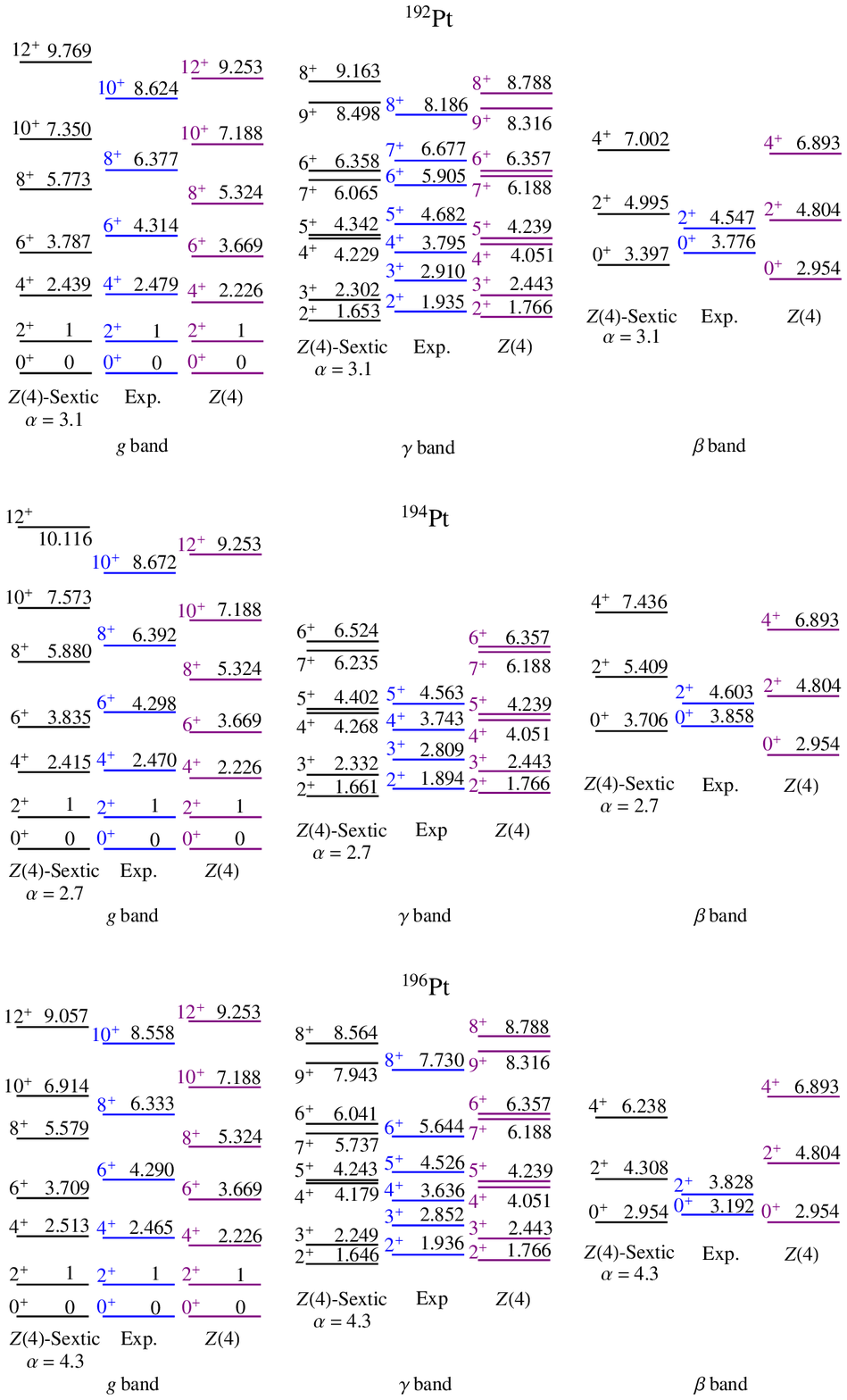}
\end{center}
\caption{The same as in Fig. \ref{fig6}, but for the experimental data \cite{Cor,Bal2,Hua} of the $^{192,194,196}$Pt isotopes. The corresponding rms values for Z(4)-Sextic are 0.614, 0.543 and 0.682, while for Z(4) one obtains 0.662, 0.707 and 0.732.}
\label{fig7}
\end{figure*}

\begin{figure}[bh!]
\begin{center}
\includegraphics[width=0.7\textwidth]{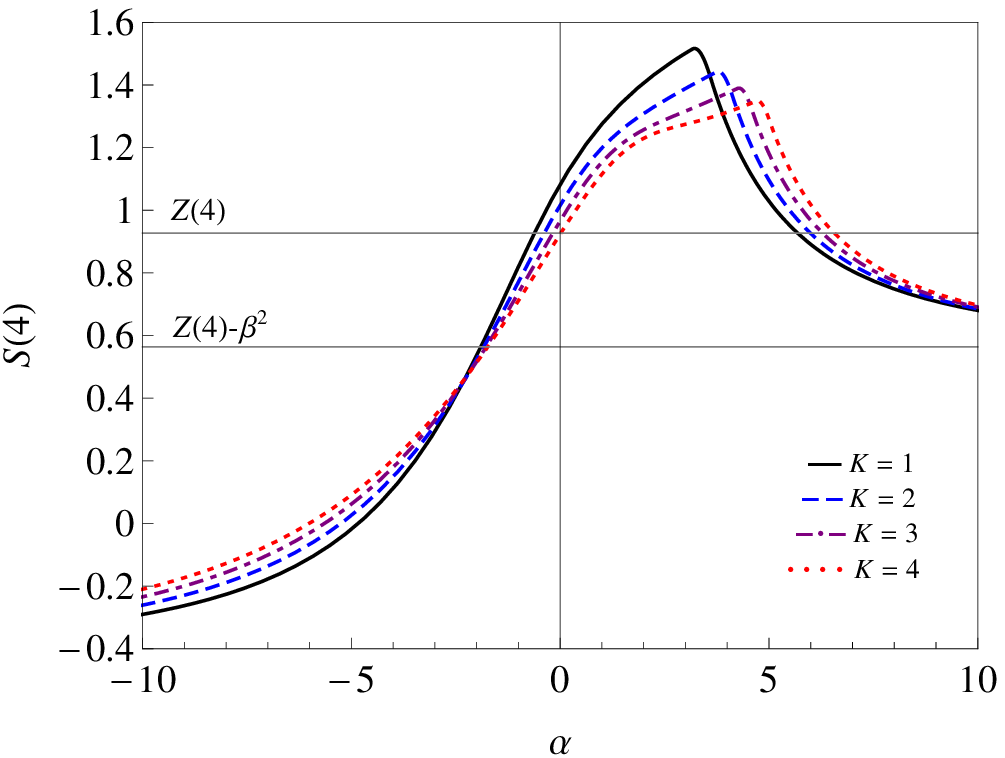}
\end{center}
\caption{The staggering $S(4)$ given by Z(4)-Sextic for $\alpha\in[-10,10]$ is compared with the values yielded by the Z(4) and Z(4)-$\beta^2$ models.}
\label{fig8}
\end{figure}

A particular signature which is often used as a characteristic of structure and its evolution is the staggering in the $\gamma$ band energies \cite{Zam} usually given in terms of the quantity $S(4)$ which is defined as:
\begin{equation}
S(J)=\frac{\left[E(L_{\gamma})-E(L_{\gamma}-1)\right]-\left[E(L_{\gamma}-1)-E(L_{\gamma}-2)\right]}{E(2_{g}^{+})},
\end{equation}
where $E$ stands for the absolute energy with respect to the ground state. In Ref.\cite{Mac} was shown that for triaxial $\gamma$ rigid cases $S(4)>0.56$, with the limiting value corresponding to Z(4)-$\beta^{2}$ model. Studying Fig. \ref{fig8}, where $S(4)$ calculated in present model is visualized as function of $\alpha$, one ascertains that the Z(4)-Sextic predictions fall in the aforementioned class for $\alpha>-2$. Moreover, comparing present calculations with the value $S(4)=0.93$ of the $Z(4)$ solution one can see that it is doubly achievable in the $\alpha>-2$ interval. In the rest of the $\alpha$ interval, $S(4)$ decreases to negative values as $\alpha\rightarrow-\infty$, but not low enough to reach the U(5)-O(6) transition region values. The highest value of $S(4)$ obviously corresponds to $\alpha_{c}$ and which is very close to that of the Davydov's triaxial rigid rotor model \cite{Dav2}. The phenomenon described above is known as the $\Delta J=1$ or even-odd staggering. Taking another look at the $\alpha$ dependent spectra of Fig. \ref{fig2}, one can observe in the ground and $\gamma$ bands another interesting phenomenon known as $\Delta J=2$ staggering or $\Delta J=4$ bifurcation which although very small was reported in the ground bands of actinide and rare earth nuclei \cite{Toki1,Toki2}. There are many theoretical approaches dedicated to this topic which are briefly mentioned in Ref.\cite{Toki2}. In the present model, this anomalous behaviour has a clear analytical origin which resides in the $\Delta L=4$ grouping of the states defined by the rules (\ref{rule1}) and (\ref{rule2}). It is interesting that the reciprocal closeness of the consecutive states is rearranged when going from negative to positive values of $\alpha$. This theoretical result hints to the fact that the $\Delta J=2$ staggering in the ground band of some nuclei can be due to higher order anharmonicities in their collective motion.

It is worth to mention that the similarities with the Z(4) and Z(4)-$\beta^{2}$ models enumerated so far reveal the fact that the approximation $\langle y^2\rangle$ used to solve the eigenvalue problem for the $\gamma$ band is a good one. The advantage of the Z(4)-Sextic, comparing with the Z(4) and Z(4)-$\beta^2$ models, is that its potential can be varied smoothly, accommodating different deformation situations and creating in this way the possibility to cover intermediate cases between the Z(4)-$\beta^2$ and $Z(4)$ or even beyond their boundaries.

\setlength{\tabcolsep}{8.5pt}
\begin{table}[ht!]
\caption{Some $B(E2)$ transitions, given by Eq. (\ref{tran}) and normalized to $B(E2;2_{g}^{+}\rightarrow0_{g}^{+})$, are compared with the experimental data \cite{Kan,Bal1,Yu} for the $^{128,130,132}$Xe isotopes and with the Z(4) model predictions. }
\begin{tabular}{ccccccccc}
\hline
\hline
$\underline{B(E2,L_{i}^{+}\rightarrow L_{f}^{+})}$&&\multicolumn{2}{c}{$^{128}$Xe}&&$^{130}$Xe&\multicolumn{2}{c}{$^{132}$Xe}&\\
\cline{3-4}\cline{7-8}
$B(E2;2_{g}^{+}\rightarrow0_{g}^{+})$&&Exp.&Z(4)-S\footnote{S is an abbreviation for Sextic. }&&Z(4)-S&Exp.&Z(4)-S&Z(4)\\
\hline
$2_{g}^{+}\rightarrow0_{g}^{+}$&&1&1&&1&1&1&1\\
$4_{g}^{+}\rightarrow2_{g}^{+}$&&1.468&1.806&&1.966&1.238&2.048&1.707\\
$6_{g}^{+}\rightarrow4_{g}^{+}$&&1.940&2.549&&2.972&&3.273&2.414\\
$2_{\gamma}^{+}\rightarrow2_{g}^{+}$&&1.194&1.771&&1.888&1.775&1.947&1.737\\
$2_{\gamma}^{+}\rightarrow0_{g}^{+}$&&0.016&0.000&&0.000&0.003&0.000&0.000\\
\hline
\hline
\end{tabular}
\label{tab1}
\end{table}

\begin{table}[ht!]
\caption{The same as in Table \ref{tab1} but for the experimental data \cite{Cor,Bal2,Hua} of the $^{192,194,196}$Pt isotopes.}
\begin{tabular}{ccccccccccc}
\hline
\hline
$\underline{B(E2,L_{i}^{+}\rightarrow L_{f}^{+})}$&&\multicolumn{2}{c}{$^{192}$Pt}&&\multicolumn{2}{c}{$^{194}$Pt}&&\multicolumn{2}{c}{$^{196}$Pt}&\\
\cline{3-4}\cline{6-7}\cline{9-10}
$B(E2;2_{g}^{+}\rightarrow0_{g}^{+})$&&Exp.&Z(4)-S\footnote{S is an abbreviation for Sextic. }&&Exp.&Z(4)-S&&Exp.&Z(4)-S&Z(4)\\
\hline
$2_{g}^{+}\rightarrow0_{g}^{+}$&&1&1&&1&1&&1&1&1\\
$4_{g}^{+}\rightarrow2_{g}^{+}$&&1.556&1.750&&1.728&1.690&&1.478&1.895&1.706\\
$6_{g}^{+}\rightarrow4_{g}^{+}$&&1.224&2.424&&1.362&2.296&&1.798&2.770&2.414\\
$8_{g}^{+}\rightarrow6_{g}^{+}$&&&3.078&&1.016&2.880&&1.921&3.622&2.913\\
$10_{g}^{+}\rightarrow8_{g}^{+}$&&&3.493&&0.691&3.238&&&4.219&3.293\\
$2_{\beta}^{+}\rightarrow0_{\beta}^{+}$&&&0.868&&&0.810&&0.123&1.047&0.769\\
$2_{\beta}^{+}\rightarrow4_{g}^{+}$&&&0.351&&&0.275&&0.003&0.590&0.422\\
$2_{\beta}^{+}\rightarrow0_{g}^{+}$&&&0.001&&&0.001&&0.000&0.000&0.005\\
$2_{\beta}^{+}\rightarrow2_{\gamma}^{+}$&&&0.099&&&0.088&&0.006&0.126&0.184\\
$0_{\beta}^{+}\rightarrow2_{g}^{+}$&&&1.362&&0.013&1.159&&0.069&1.852&1.151\\
$0_{\beta}^{+}\rightarrow2_{\gamma}^{+}$&&&0.000&&0.171&0.000&&0.443&0.000&0.000\\
$6_{\gamma}^{+}\rightarrow4_{\gamma}^{+}$&&&1.028&&&0.974&&1.207&1.175&1.142\\
$4_{\gamma}^{+}\rightarrow2_{\gamma}^{+}$&&&0.750&&0.427&0.724&&0.714&0.812&0.801\\
$3_{\gamma}^{+}\rightarrow2_{\gamma}^{+}$&&1.783&2.251&&&2.183&&&2.415&2.365\\
$6_{\gamma}^{+}\rightarrow6_{g}^{+}$&&&0.224&&&0.211&&0.394&0.262&0.218\\
$6_{\gamma}^{+}\rightarrow4_{g}^{+}$&&&0.000&&&0.000&&0.012&0.000&0.000\\
$4_{\gamma}^{+}\rightarrow4_{g}^{+}$&&&0.365&&0.285&0.347&&&0.415&0.381\\
$4_{\gamma}^{+}\rightarrow2_{g}^{+}$&&&0.000&&0.007&0.000&&0.014&0.000&0.000\\
$3_{\gamma}^{+}\rightarrow4_{g}^{+}$&&0.664&1.339&&&1.280&&&1.489&1.360\\
$3_{\gamma}^{+}\rightarrow2_{g}^{+}$&&0.012&0.000&&&0.000&&&0.000&0.000\\
$2_{\gamma}^{+}\rightarrow2_{g}^{+}$&&&1.730&&1.809&1.684&&&1.837&1.737\\
$2_{\gamma}^{+}\rightarrow0_{g}^{+}$&&0.010&0.000&&0.006&0.000&&&0.000&0.000\\
\hline
\hline
\end{tabular}
\label{tab2}
\end{table}

\begin{figure}[bh!]
\begin{center}
\includegraphics[width=0.7\textwidth]{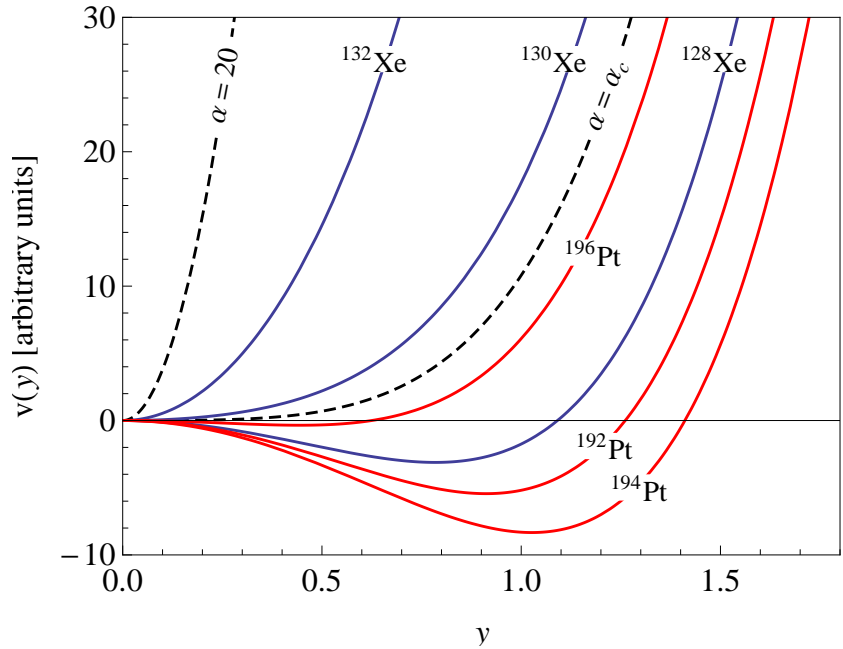}
\end{center}
\caption{Ground state potential (\ref{poty}) with $K=4$ plotted as function of $y$ for all considered nuclei with $\alpha$ resulting from the fits of Figs. \ref{fig6} and \ref{fig7}. The critical point potential ($\alpha=\alpha_{c}$) as well as a potential close to the spherical limit ($\alpha=20$) are given for reference.}
\label{Fig9}
\end{figure}

As in Ref.\cite{Bona2}, Z(4)-Sextic is applied in Fig. \ref{fig6} for $^{128,130,132}$Xe isotopes which were considered as candidates for Z(4) model and additionally in Fig. \ref{fig7} for $^{192,194,196}$Pt isotopes. The best fits of the experimental energy spectra were obtained for $K=4$, but not very different from $K=2$ and $3$, while the $K=1$ case obviously has only theoretical importance. The experimental data of these nuclei are better described by Z(4)-Sextic comparing to Z(4), except for the $^{130}$Xe nucleus as can be deduced from the corresponding rms values given in the captions of Figs. \ref{fig6} and \ref{fig7}. The ground and $\gamma$ bands of the Pt isotopes are similarly described by both models, while the $\beta$ band is much better reproduced by Z(4)-Sextic. The poor description of the $\beta$ band within $Z(4)$ model is the reason why these nuclei were never considered as Z(4) candidates. In what concerns the Xe isotopes, their description by means of the present approach proved to be better than that of Z(4) only for $^{128}$Xe and $^{132}$Xe, although for $^{130}$Xe both models give very similar results. It is interesting that while the $\gamma$ band of $^{128}$Xe has a similar description in both models, the ground and $\beta$ bands are slightly better described within present approach. The picture is quite different in case of $^{132}$Xe where one obtains the biggest difference between the rms values, in favor of Z(4)-Sextic fit. Indeed, even if Z(4) describes the $\beta$ band better it fails to do the same for ground and $\gamma$ bands. Moreover, our model simulates very closely the $\Delta J=2$ staggering of the experimental ground band. The overall impression of the comparison with experiment is that the ground and $\beta$ bands are quite well described for all considered nuclei, while the specific $\gamma$ band staggering of the rigid triaxial models in general, is found only in the experimental spectrum of $^{194}$Pt. Plotting in Fig. \ref{Fig9} the potential (\ref{poty}) using the fitted values of $\alpha$, one observes the following features. The advantage of Z(4)-Sextic over the pure oscillator or infinite square well potentials, is that it can describe a shape phase transition. This can be clearly seen from the shapes of the potentials resulted for Xe isotopes, where $^{128}$Xe has a potential with a deformed minimum, $^{132}$Xe with a spherical minimum while $^{130}$Xe is situated in the critical point region. This result confirms $^{130}$Xe as being a good candidate for the Z(4) model. Another important point emerging from Fig. \ref{Fig9} is that using a sextic potential with a deformed minimum one obtained a relatively good agreement with experiment for Pt isotopes, which could not be achieved using simpler potentials such as harmonic oscillator and infinite square well.

Concerning  the $B(E2)$ transitions, the agreement of Z(4)-Sextic is very good with all the available data for the $^{128,130,132}$Xe isotopes. In what concerns the $^{192,194,196}$Pt isotopes, in the ground band the Z(4)-Sextic and Z(4) numerical results provide a good agrement with experiment only for the $4_{g}^{+}\rightarrow2_{g}^{+}$ transition, the rest of the transition probabilities being overestimated in both calculations. A possible way to improve the agreement is to add anharmonicities to the transition operator as was made in Ref. \cite{Raduta}. For transitions in the $\gamma$ band and from the $\gamma$ band to the ground band both approaches give good results, while for transitions from the $\beta$ band to the ground and $\gamma$ bands the agreement is only partially good. These applications show that these isotopes can be considered partial candidates for Z(4)-Sextic and Z(4). The good agreement for all three bands of the isotope $^{194}$Pt proves that these solutions can describe real situations and opens the question if there are better candidates.

\section{Conclusions}

The main result of the present work consists in the proposal of a new solution for the Davydov-Chaban Hamiltonian, with a sextic oscillator potential for the variable $\beta$ and $\gamma$ "frozen" to $30^{\circ}$. The solution is conventionally called Z(4)-Sextic, in connection with the precedent Z(4) solution where an infinite square well potential was considered. Choosing a quasi-exactly solvable form for the sextic potential, a finite set of states was analytically determined. The corresponding eigenvalue problem is exactly solved in the case of ground and $\beta$ bands, while for the $\gamma$ band states an approximation is involved. The difference from the former quasi-exactly solvable sextic potential approaches \cite{Lev1,Lev2,Rad1,Rad2}, is the introduction of a completely different scheme for angular momentum attribution which satisfy the condition of constant potential. Also it is the first time when the scaling property of the problem is employed to describe the properties of the quasi-exactly solvable sextic potential. Indeed, as was shown in Section III, the model depends up to a scaling factor on a single parameter. Taking advantage of this property, one studied the evolution of the energy spectra and the corresponding transition probabilities when the free parameter is varied through different shapes of the associated sextic potential. For two values of the free parameter, the potential has one vanishing term. The spectra normalized to the energy of the first exited state and the $B(E2)$ transitions normalized to the transition between the first excited state and the ground state calculated with the present model for these special cases, constitute parameter independent realizations of the associated simplified sextic potentials.

A detailed comparison to the Z(4) and Z(4)-$\beta^{2}$ models, especially in terms of the energy spectrum, revealed that the present formalism approximate quite well the former in its critical point, and exactly reproduces the latter in the asymptotic limit of the free parameter. These facts suggest the consistency of the approximation used for the treatment of the $\gamma$ band states.

Numerical applications were performed for $^{128,130,132}$Xe and $^{192,194,196}$Pt isotopes. The results for the Xe isotopes revealed a shape phase transition with its critical point identified with the $^{130}$Xe nucleus. The fits have a qualitative character, showing that the experimental realization of triaxial $\gamma$ rigidity is very much possible. Especially encouraging in this sense is the reproduction of the $^{194}$Pt spectrum.

Concluding, one should say that the theoretical value of the proposed model resides in the fact that it adds to the few exactly solvable solutions of the collective model concerning only the ground and $\beta$ bands, while its special cases contribute to the even more restrained set of parameter free models.

\begin{acknowledgments}
This work was supported by the Romanian Ministry for
Education Research Youth and Sport through the CNCSIS
Project ID-2/5.10.2011.
\end{acknowledgments}

\end{document}